\documentclass[preprint,aps,epsfig]{revtex4}
\usepackage[dvips]{graphicx}
\begin{document}

\tighten
\draft
\preprint{
\vbox{
\hbox{\today}
\hbox{Tashkent}
}}

\newcommand{\re}[1]{(\ref{#1})}
\newcommand{\lab}[1]{\label{#1}}
\newcommand{\ci}[1]{\cite{#1}}
\renewcommand{\baselinestretch}{1.25}
\newcommand{\bfr}{\begin{flushright}}
\newcommand{\bfl}{\begin{flushleft}}
\newcommand{\efl}{\end{flushleft}}
\newcommand{\efr}{\end{flushright}}
\newcommand{\bc}{\begin{center}}
\newcommand{\ec}{\end{center}}
\newcommand{\be}{\begin{equation}}
\newcommand{\ee}{\end{equation}}
\newcommand{\bea}{\begin{eqnarray}}
\newcommand{\eea}{\end{eqnarray}}
\newcommand{\ba}{\begin{array}}
\newcommand{\ea}{\end{array}}
\newcommand{\edc}{\end{document}}
\newcommand{\ul}{\underline}
\newcommand{\ri}{\rightarrow\infty}
\newcommand{\li}{\leftarrow\infty}
\newcommand{\ra}{\rightarrow}
\newcommand{\la}{\leftarrow}
\newcommand{\ds}{\displaystyle}
\newcommand{\dsf}{\displaystyle\frac}
\newcommand{\dt}{\Delta{t}}
\newcommand{\il}{\int\limits}
\newcommand{\pal}{\partial}
\newcommand{\xxx}{{\it{X}}}
\newcommand{\bone}{{\bf 1}}
\newcommand{\gComment}[1]{}
\renewcommand{\gComment}[1]{\textcolor{red}{Gerardo: #1}}

\title{
Analytic approach to the ground-state energy of charged anyon
gases }


\author{B. Abdullaev$^1$, U. R\"{o}ssler$^2$,
and M. Musakhanov$^1$ }

\address{$^1$
Department of Theoretical Physics, Institute of Applied Physics,
Uzbekistan National University,
 Tashkent 100174, Uzbekistan}
\address{$^2$
Institute for Theoretical Physics, University of Regensburg, D-93040
Regensburg, Germany}

\date{Received \today }

\begin{abstract}
We derive an approximate analytic formula for the ground-state
energy of the charged anyon gas. Our approach is based on the
harmonically confined two-dimensional (2$D$) Coulomb anyon gas and
a regularization procedure for vanishing confinement. To take into
account  the fractional statistics and Coulomb interaction we
introduce a function, which  depends on both the statistics and
density parameters ($\nu$ and $r_s$, respectively). We determine
this function by fitting to the ground state energies of the
classical electron crystal at very large $r_s$ (the 2$D$ Wigner
crystal), and to the Hartree-Fock (HF) energy of the
spin-polarized 2$D$ electron gas, and the dense 2$D$ Coulomb Bose
gas at very small $r_s$. The latter is calculated by use of the
Bogoliubov approximation. Applied to the boson system ($\nu=0$)
our results are very close to recent results from Monte Carlo (MC)
calculations. For spin-polarized electron systems ($\nu=1$) our
comparison leads to a critical judgment concerning the density
range, to which the HF approximation and MC simulations apply.
In dependence on $\nu$, our analytic formula yields ground-state energies, which monotonously
increase from the bosonic to the fermionic side if $r_s > 1$. For $r_s\leq 1$
it shows a nonmonotonous behavior indicating a breakdown of the assumed
continuous transformation of bosons into fermions by variation of the parameter
$\nu$.
\end{abstract}

\maketitle


\newpage

\section{Introduction}
\label{sec1}

The jellium model \ci{Kittel63}, consisting of a homogeneous
electron (or fermion) gas charge-neutralized by a uniform positive
background, is serving as a realistic model to describe generic
properties of metallic systems, which extend in two (2$D$) or
three (3$D$) spatial dimensions \ci{Ginliani}. Especially, the
exchange and correlation contributions to the ground state energy
per particle in dependence on the particle density (quantified by
the Wigner-Seitz radius $r_s$) and the degree of spin polarization
are required in applications of the density functional theory (in
its local density approximation) to inhomogeneous fermion systems
({\it e.g.}, nuclei, molecules, solids) \ci{Dreizler90}. Analytic
formulas of these contributions for 2$D$ and 3$D$ systems are
available from interpolations between known data from numerical or
analytical calculations for special values of the system
parameters \ci{attac,tanatar,seidl}. Although frequently in use,
such formulas are not free of shortcomings and may lead to
misleading results \ci{Wensauer}. Here, the special topology of 2D
systems provides a unique opportunity to derive an approximate
analytic formula for the ground state energy, which has not been
exploited so far.

The 2D topology allows for fractional exchange statistics \ci{lei}, characterized
by a continuous parameter $\nu$, that may attain values between 0 (for bosons)
and 1 (for fermions). Particles with $0<\nu<1$ are generically called anyons
\ci{wil2,wil1}. The quasiparticle excitations in the fractional quantum Hall regime
\ci{fis,ler,kh} and in certain quantum magnets \cite{kitaev} can be
described using the anyon concept.

2$D$ electron systems are realized in semiconductor
heterostructures and layered metallic systems such as cuprate
superconductors. Their ground-state properties are subjects of
fundamental studies (see Ref. \ci{kwon} and references therein).
To obtain accurate estimates of the ground-state energy, numerical
simulations have been employed \ci{attac,tanatar}, which are
restricted, however, to special values of the particle density
(expressed by the interparticle distance $r_s$) or the degree of
spin polarization. An analytic expression in terms of $r_s$ exists
for the 2$D$ Wigner crystal ($r_s\gg 1$) \ci{bm}, which represents
a classical  system. All these investigations refer to 2$D$
electron systems and are, as such, based on the fermion character
of the particles. It is interesting to note that bosonic Coulomb
systems have found so far only little attention \ci{DePalo}, which may
be due to lacking realization. This seems to be even more the case
for the anyon aspects of the 2$D$ Coulomb gases, while for the
3$D$ Coulomb Bose gas a closed-form expression for the ground-state 
energy as a function of $r_s$  was derived by use of the
Bogoliubov approximation \ci{foldy}. The aim of this paper is, by
making use of the anyon concept, to establish a link between the
analytic results, which exist for the ground-state energies of the
2$D$ Coulomb Bose gas (derived here in close analogy to Ref.
\ci{foldy}), of the Wigner crystal \ci{bm}, and of the
high-density 2$D$ spin-polarized electron gas.

Previously, we have derived an approximate analytic expression for the
ground state energy of $N$ charged anyons confined in a 2$D$ harmonic
potential  \ci{aormn}. This was achieved by using the bosonic
representation of anyons and a gauge vector potential to account for
the fractional statistics, which allowed us to work with the product
ansatz for the $N$-body wave function. A variational principle has been
applied  by constructing this wave function from single-particle
Gaussians of variable shape. As in many other perturbative treatments
of anyons in an oscillator potential (see Ref. \ci{aormn} and
references therein) our expression for the ground state energy had a
logarithmic  divergence connected with a cutoff parameter for the
interparticle distance. Making use of the physical argument (see
Ref.~\ci{ler}) that for $\nu\neq 0$  this distance has to have some
finite value, we have regularized the formula obtained for the ground
state energy by an appropriate procedure that takes into  account the
numerical results for electrons in quantum dots in the case with
Coulomb interaction.

The ground state of anyons in a harmonic potential has been studied by Chitra and Sen \ci{chitra},
especially for the limit of an infinite number of particles, yet without Coulomb
interaction. In the present paper, we make use of previous work for the harmonically
confined 2D anyons \ci{aormn, chitra} to derive an approximate analytic
expression for the ground-state energy of the homogeneous 2D anyon gas. This is
done (for all values of the statistic parameter $\nu$) by flattening out the
confining potential with a simultaneous increase of the particle number $N$, but
fixed areal density, to obtain the infinite size system, i.e., the thermodynamic
limit. It is achieved by redefining the strength $\omega_0$ of the harmonic
potential such that it vanishes with increasing $N$ as $1/N^{1/2}$. With respect
to the dependence on the statistic parameter we have to distinguish between the
cases without and with Coulomb interaction. In the former case, we introduce a
function of $\nu$, which is fitted to known analytic expressions at the fermion
or boson end. For the case with Coulomb interaction, one has to warrant also
charge neutrality. This is achieved by introducing the interaction with a
positively charged background disk, which slightly modifies the interaction term
in the formula for the ground-state energy. For this case, as the dependence on
$\nu$ may now be mixed with that on the density parameter $r_s$, we assume a
function, which depends on both parameters. This function is fitted to the
analytic expressions for the ground state energies of the
classical Wigner crystal (for large $r_s$ and independent of
$\nu$) and of the 2$D$ Coulomb Bose gas (for small $r_s$ and
$\nu=0$). For the fermion case ($\nu=1$), we fit this function to
the analytic expression for the HF energy, which is a high-density
limit ($r_s\rightarrow 0$).

In Sec. \ref{sec2} we introduce the principal approach for
deriving an analytic formula for the ground state energy and
demonstrate the regularization to obtain the thermodynamic limit.
In Sec. \ref{sec3} we derive an analytic expression for the
ground-state energy of the 2$D$ Coulomb Bose gas using the
Bogoliubov approximation. In Secs \ref{sec4} and \ref{sec5}
the derivation of the approximate analytic formula for the ground
state energy of the 2$D$ Coulomb anyon gas is completed and the
results will be discussed by comparing with numerical results from
the literature. We summarize and conclude with Sec. \ref{sec6}.

\section{Principal approach and harmonic potential regularization}
\label{sec2}

Let us start by considering $N$ noninteracting spinless anyons of mass
$M$  confined in a $2D$ parabolic potential described by the
Hamiltonian
\be
\hat H_0=\dsf{1}{2M}\ds\sum_{i=1}^N\left[\left(\vec p_i+\vec
A_{\nu}(\vec r_i)\right)^2+M^2\omega_0^2 |\vec{r_i}|^2\right] \, .
\lab{gsetup1}
\ee
Here $\vec r_i$ and $\vec p_i$ are the position and momentum,
respectively, of the $i$th anyon in 2$D$ and
\be
\vec A_{\nu }(\vec r_i)=\hbar\nu\ds\sum_{j\not=i}^N\dsf{\vec e_z
\times\vec r_{ij}} {|\vec r_{ij}|^2}
\lab{gsetup2}
\ee
is the anyon gauge vector potential \ci{wu,lau} with  $\vec r_{ij}=\vec
r_i-\vec r_j$, and $\vec e_z$  the unit vector normal to the 2$D$
plane. The parameter  $\nu$ accounts for the fractional statistics  of
the  anyon: it varies between $\nu=0$  for bosons and $\nu=1$ for
fermions. When later on considering the anyons as charged particles, we
add their mutual interaction to define the Coulomb anyon problem.

In Ref. \ci{aormn} we have outlined a variational procedure for the
ground-state energy of interacting anyons confined in an oscillator
potential with characteristic energy $\hbar\omega_0$. Starting from
the bosonic end ($\nu = 0$) we achieved, after regularization of a
logarithmic expression by means of a cutoff parameter for the
particle-particle interaction, approximate analytic formulas in
terms of $N$ and $\nu$. For the noninteracting anyon system we
found \be E_0(N, \nu )=\hbar \omega_0 N {\cal N}^{1/2} \, ,
\lab{gsetup3} \ee while for the interacting anyon system  it is
given by \be E_0 (N, \nu )=\dsf{\hbar \omega_0 N}{2}\left[\dsf{\cal
N}{X_0^2}+X_0^2+ \dsf{2{\cal M}}{X_0} \right] \, , \lab{gsetup4} \ee
with \be X_0=(A+B)^{1/2}+[-(A+B)+2(A^2-AB+B^2)^{1/2}]^{1/2} \, ,
\lab{gsetup5} \ee and \bea \ba{l} A=\left[{\cal
M}^2/128+\left(({\cal N}/12)^3+({\cal M}^2/128)^2
\right)^{1/2} \right]^{1/3},\\
B=\left[{\cal M}^2/128-\left(({\cal N}/12)^3+({\cal M}^2/128)^2
\right)^{1/2} \right]^{1/3} \, .
\lab{gsetup6}
\ea
\eea
In these expressions we use ${\cal N}=1+\nu(N-1)$ and
\bea
\ba{l}
{\cal M}=\left(\dsf{\pi}{2}\right)^{1/2}
\dsf{N-1}{2}\dsf{L}{a_B} \, ,
\lab{gsetup7}
\ea
\eea
where $L=(\hbar /M\omega_0)^{1/2}$ is the oscillator unit length, and
$a_B$  the Bohr radius.

In order to obtain the corresponding expressions for the
homogeneous 2$D$ anyon gas, we flatten out the parabolic confining
potential while increasing  the number $N$ of anyons, but keeping
the density $\rho = N/S=1/\pi r_0^2$ constant, i.e.  we perform
the thermodynamic limit while making the confining potential
disappear. Here $ \pi r_0^2$ is the area of the jellium disk
carrying the positive countercharge corresponding to the mean
particle distance $r_0=a_B r_s$ expressed in units of the Bohr
radius by the dimensionless density parameter $r_s$.

Without Coulomb interaction and in the case of fermions ($\nu=1$),
the ground state energy of the homogeneous 2$D$ electron system of
density $\rho $ is determined by the Pauli exclusion principle. It
is given by \be E_0(\rho)=\pi \hbar^2 \rho N/M \, , \lab{gsetup8}
\ee while from Eq. \re{gsetup3} we have (see also Ref.
\ci{chitra} for $\nu=1$ and $N\rightarrow \infty$) \be E_0(N,
\nu=1 )=\hbar \omega_0 N^{3/2} \, . \lab{gsetup9} \ee In the
thermodynamic limit both expressions have to become identical and
we obtain the relation \be \omega_0(N)= \pi \hbar \rho /(M
N^{1/2}) \, , \lab{gsetup10} \ee which means that, in fact, the
thermodynamic limit ($N\rightarrow \infty$) is obtained for the
vanishing parabolic confining potential. We extend this
consideration for the fermionic limit, $\nu=1$, to the general
anyon case,  $\nu \neq 1$, by assuming instead of Eq. \re{gsetup8} the
relation \be E_0(\rho, \nu)=\pi \hbar^2 \rho N \phi(\nu)/M \, ,
\lab{gsetup11} \ee where the function $\phi(\nu)$ is still to be
determined under the constraint  $\phi(\nu=1)=1$. This form is
motivated by the fact that close to the bosonic  limit ($\nu
\simeq 0$) the ground-state energy of the infinite anyon gas
depends  linearly on $\nu$ \ci{sen,wen,mori}. We replace this
particular dependence by $\phi(\nu)$. Consequently, we have to
change Eq. \re{gsetup10} into \be \omega_0(N, \nu)= \pi \hbar \rho
f(\nu) /(M N^{1/2}) \lab{gsetup12} \ee with another unknown
function $f(\nu)$  and the constraint $f(\nu=1)=1$. Thus we
assume the vanishing of $\omega_0$  according to $1/ N^{1/2}$ also
in the general anyon case $\nu \not= 1$. As it turns out
$\phi(\nu)$  is determined  by $f(\nu)$. For the free
(noninteracting) anyon gas both unknown functions would be the
same. However, in our generalization to the charged anyon gas this
will not be the case anymore.

In the thermodynamic limit, and including the Coulomb interaction,
the parabolic confinement (caused by geometry and electrostatics
of the compensating charges) has to be replaced by the jellium
contribution, which for a disk of radius $R_0$  (containing $N$
countercharges) gives a potential energy contribution \ci{laupr}
\be V(\vec r_k)=-\rho \int_{S} \dsf{ e^2 \ d^2 r}{|\vec r_{k}-\vec
r|} \lab{gsetup13} \, . \ee Here $S=\pi R_0^2$ is the area of the
jellium disk for $N$ charges and we have  $R_0=N^{1/2} r_0$. For
$\nu = 1$, we may now identify the characteristic length  $L$ of
the oscillator with the mean particle distance to obtain $L=(\hbar
/M\omega_0)^{1/2}=N^{1/4}r_0$, and find for $N\gg 1$ the relation
$r_0\ll L\ll R_0$. The generalization to the anyon case, $\nu
\not= 1$, is possible by dividing those lengths by $f^{1/2}(\nu)$.

The strength of the particle interaction is characterized by the
density parameter $r_s$ (in 2$D$ the radius of the Seitz circle),
and we may generalize our analytic expression \re{gsetup11} for
the ground state-energy to the form \be E_0(\nu, r_s)=\pi \hbar^2
\rho N \phi(\nu, r_s)/M \, . \lab{gsetup14} \ee

Thus we replace the unknown function $\phi(\nu)$ by $\phi(\nu,
r_s)$, which now depends on the two system parameters $\nu$ and
$r_s$ and takes into account the Coulomb interaction and the
statistics. In the high density limit $(r_s\rightarrow 0)$ it
becomes the function $\phi(\nu)$ introduced for the
noninteracting system. The nontrivial case of $r_s \rightarrow
0$ and $\nu \rightarrow 0$ will be considered in Sec.
\ref{sec5}.

 In a similar way, we have to generalize $f(\nu)$ to
$f(\nu, r_s)$. Now these functions depend in a complicated way on
each other. Thus we have established the extension from the
confined to the homogeneous 2$D$ anyon system. In Sec.
\ref{sec5} we determine the function $f(\nu, r_s)$.

\section{2$D$ Coulomb Bose gas at high densities}
\label{sec3}

The 3$D$ Coulomb Bose gas problem has been treated by Foldy
\ci{foldy} in the high density limit by applying the Bogoliubov
approximation \ci{bogoliubov}. Originally, Bogoliubov considered a
low-density system of bosons interacting with short-range forces,
but it was shown in Ref. \ci{foldy} that the Bogoliubov
approximation can also be reliably used for bosons with long-range
Coulomb interactions in the high density limit. In this section we
use the same strategy for the 2$D$ Coulomb Bose gas to obtain an
analytic expression for the ground-state energy.

According to Ref. \ci{foldy}, the Bogoliubov approximation is valid if
almost all particles are in the zero momentum state, i.e.,
$(N-N_0)/N$ tends to zero when  $r_s \rightarrow 0$, where $N_0$ is
the number of particles with zero momentum. Following Ref. \ci{foldy},
for the 2$D$ Coulomb Bose gas, this ratio takes the form \be
\dsf{(N-N_0)}{N}=\dsf{S}{4\pi N}\int_{0}^{\infty} \left(\dsf{{\cal
E}_k+NV_k}{E_k}-1\right) k \ d k \, , \lab{2dbg1} \ee where ${\cal
E}_k=\hbar^2k^2/(2M)$, $V_k=2\pi e^2/(Sk)$ is the Fourier transform
of the $1/r$ potential in 2$D$, and \be E_k=[({\cal
E}_k+NV_k)^2-N^2V_k^2]^{1/2} \lab{2dbg2} \ee is the dispersion of
collective excitations. By introducing the variable \be
\xi=\left(\dsf{\hbar^2 k^{3}S}{2 \pi M N e^2} \right)^{1/6} =
\left(\dsf{a_B^3 r_s^2 k^{3} }{2} \right)^{1/6} \, , \lab{2dbg3} \ee
Eq. \re{2dbg1} takes the form \be
\dsf{(N-N_0)}{N}=\dsf{r_s^{2/3}}{2^{1/3}}\int_{0}^{\infty}
\left[\dsf{\xi^6+2}{(\xi^6+4)^{1/2}}- \xi^3\right] \ d\xi \, .
\lab{2dbg4} \ee After evaluation of the integral one obtains \be
\dsf{(N-N_0)}{N}=\frac{\Gamma(\frac{1}{3})\Gamma(\frac{7}{6})}{2
\sqrt{\pi}}r_s^{2/3}=0.701091  r_s^{2/3} \, , \lab{2dbg5} \ee which
tends to zero for $r_s \rightarrow 0$, thus showing the validity of
the Bogoliubov  approximation also for the 2$D$ Coulomb  Bose gas.

The 2$D$  analog of the ground-state energy given in Ref. \ci{foldy} is
\be E=\dsf{S}{4\pi}\int_{0}^{\infty}(E_k-{\cal E}_k-NV_k) k \ d k \,
, \lab{2dbg6} \ee and can also be written in terms of the variable
$\xi$ (in Ry=$e^2/(2a_B)$ units) \be
\dsf{E}{N}=2^{1/3}r_s^{-2/3}\int_{0}^{\infty}
[\xi^3(\xi^6+4)^{1/2}-\xi^6-2]\xi \ d\xi \, , \lab{2dbg7} \ee which
after evaluation of the integral yields \be \dsf{E}{N}=-
c_{BG}r_s^{-2/3} \, , \lab{2dbg8} \ee where
$c_{BG}=\frac{2\Gamma(-\frac{4}{3})\Gamma(\frac{5}{6})}{3
\sqrt{\pi}}=1.29355$. The same result has been obtained also in Ref.
\ci{Apaja} by applying the hypernetted chain approximation.

Thus we have an exact analytic expression for
the ground-state energy per particle for the 2$D$ Coulomb Bose gas
valid at high densities. It will be used in Sec. \ref{sec5} together
with the known expression for the 2$D$ Coulomb gas in the low density
limit (the 2$D$ Wigner crystal) to derive a form for the unknown
function $f(\nu, r_s)$ introduced in Sec. \ref{sec2}. We note in
passing that the 2$D$ (classical) Wigner crystal is independent
of particle statistics, thus its properties do not depend
on $\nu$.

It is interesting to note that the spectrum of collective
excitations of the 2$D$ Coulomb Bose gas, Eq. \re{2dbg2}, can be
cast in the form \be E_k=[\hbar^2 2\pi e^2\rho
k/M+(\hbar^2k^2/(2M))^2]^{1/2} \, , \lab{2dbg9} \ee which for small
$k$ represents the 2$D$ plasmon dispersion \ci{stern} and
approaches for large $k$ the free particle dispersion. In contrast
to the 3$D$ system this spectrum has no gap at $k=0$.

\section{Analytic expression for the ground state-energy in
the thermodynamic limit} \label{sec4}

With the considerations of the previous sections we may now derive
the wanted analytic expression for the ground-state energy of the
charged anyon gas. The system Hamiltonian is given by \be \hat
H=\ds\sum_{i=1}^N\left[ \dsf{1}{2M}\left(\vec p_i+\vec
A_{\nu}(\vec r_i)\right)^2+ \dsf{1}{2}\left( \ds\sum_{j\not=i}
\dsf{e^2} {|\vec r_i - \vec r_j|}+V(\vec r_i)\right) \right] \, ,
\lab{gsqa1} \ee where (for $N\rightarrow \infty$) $V(\vec r_i)$,
the interaction energy of the $i$th electron with the jellium
background, plays the role of the confining potential considered
in Ref. \ci{aormn}.

From our calculation for parabolically confined interacting anyons
we know the contributions of the individual terms of Eq.
\re{gsqa1} to the expectation value of the ground state energy
except for the potential energy $V(\vec r_i)$. As will be
demonstrated in this section, $V(\vec r_i)$ modifies the
expectation value originating from the Coulomb interaction in the case
of $N\rightarrow \infty$. Namely, we will show that, differing
from Eq. \re{gsetup7}, ${\cal M}$ will now be proportional to
$N^{1/2}$. This allows us to obtain in Sec. \ref{sec5} the
thermodynamic limit of the ground-state energy, Eq. \re{gsetup4},
as a complicated function of the system parameters  $\nu $ and
$r_s$ introduced by ${\cal M}$ and its dependence on $f(\nu,r_s)$.

Before this is done, we have to calculate the mentioned
modification of the contribution of the Coulomb interaction to the
ground-state energy by $V(\vec r_i)$. For this we adopt the
approach of Fisher {\it et al.} \ci{fhm} who used the Ewald
method.  They started dividing the plane into square cells, each
cell containing the same number of pointlike charges (electrons)
and the corresponding jellium background. We use their expression
for the interaction potential (Eq. (A2) of Ref. \ci{fhm})

\be \Phi (\vec r) = \dsf{2\pi}{{\cal L}^2} \sum_{\vec S \not={\bf
0}} \dsf{{\sf erfc} (S t)e^{i \vec S \cdot \vec r}}{S}+ \sum_{{\cal
\vec R}} \dsf{{\sf erfc} (|\vec r-{\cal \vec R}|/2t)}{|\vec r-{\cal
\vec R}|} - \dsf{4 \pi^{1/2} t}{{\cal L}^2} \, .
\lab{ap1}
\ee
Here,
${\cal \vec R}={\cal L}(l_x,l_y)$, and $\vec S=2\pi (l_x,l_y)/{\cal
L}$ (with $l_x,l_y=0,\pm 1,\pm 2,...$) are the direct and reciprocal
lattice vectors, respectively, with lattice constant ${\cal L}$;
${\sf erfc} (x)=1-{\sf erf}(x)$ is the complementary error function,
and $t$ the Ewald  parameter for optimal convergence of the lattice
sums.

The interaction energy of the 2$D$ Coulomb gas is given by
\be
V(\vec{r}_1,...\vec{r}_N)=\dsf{1}{2} \sum_{i,j=1}^N
\left[ \Phi (\vec r_i-\vec r_j)-
\delta _{ij} \dsf{e^2}{|\vec r_i-\vec r_j|} \right] \, ,
\lab{ap2}
\ee
where the self-interaction is explicitly subtracted. Using the Gaussian
variational wave function of Ref. \ci{aormn} to compute the expectation
value of the interaction energy, we need to evaluate
\be
\sum_{i,j\not=i} \int  \ d^2 r_i \ d^2 r_j
e^{-\alpha \vec r_i^2} \ e^{-\alpha \vec r_j^2}\Phi(\vec r_i-\vec r_j) \, .
\lab{ap3}
\ee

Consider now the individual contributions to  $\Phi(\vec r_i-\vec
r_j)$ in Eq. \re{ap1}. Typically, the Ewald parameter is taken to
be $t\approx {\cal L}$. For simplicity we assume  $t= {\cal L}$.
Then $S t=2\pi (l_x^2+l_y^2)^{1/2}$, $|\vec r-{\cal \vec R}|={\cal
L}|\vec r/{\cal L}- (l_x,l_y)|$, and  $|\vec r-{\cal \vec
R}|/(2t)=|\vec r/{\cal L}-(l_x,l_y)|/2$ and we rewrite the
expression for $\Phi(\vec r)$, Eq. \re{ap1}, in the form \be
\Phi(\vec r)=\dsf{{\sf erfc}(r/2{\cal L})}{r}+ \dsf{1}{{\cal
L}}{\cal A}(\vec r) \, , \lab{ap4} \ee where \be {\cal A}(\vec
r)=\sum_{l_x, l_y \not=0} \left[\dsf{{\sf erfc} (2\pi
(l_x^2+l_y^2)^{1/2})e^{i \vec S \cdot \vec
r}}{(l_x^2+l_y^2)^{1/2}}+ \dsf{{\sf erfc} (|\vec r/{\cal
L}-(l_x,l_y)|/2)} {|\vec r/{\cal L}-(l_x,l_y)|} \right]- 4
\pi^{1/2} \, . \lab{ap5} \ee

The length of the vector $\vec{r}$ is in the interval
$0<r\leq\cal{L}$. We identify ${\cal L}\equiv L= N^{1/4} r_0$, which
implies  $r_0\leq r\leq N^{1/4} r_0$, and estimate the contributions
to  $\Phi(\vec r)$ in the limits $r\ll L$ and $r\rightarrow L$. For
$r/L\ll  1$ one has $e^{i \vec S \cdot \vec r}\approx 1$ and one can
neglect $\vec r/{\cal L}$ in the second term of ${\cal A}(\vec r)$
to obtain the leading terms  of an expansion $\Phi(\vec r)\approx
1/r+A/L$.  When $r\rightarrow L$ and with $\vec{S}\cdot\vec{r} =
2\pi(l_x^2 + l_y^2)^{1/2}\cos{\beta}$ , we find that the first term
of  ${\cal A }(\vec{r})$ depends only on $l_x, l_y$ and the angle
between   $\vec S$ and $\vec r$. Likewise, it is seen that the
second term of  ${\cal A}(\vec{r})$ neither depends on $r$ nor
on $L$. Hence we  conclude with the relation \be
\dsf{\Phi(r\rightarrow r_0)}{\Phi(r\rightarrow L)} \sim N^{1/4} \, ,
\lab{ap6} \ee showing that in the thermodynamic limit ($N\rightarrow
\infty$) the small $r$ contributions to the interaction energy
become essential.

Thus, assuming $\Phi(\vec r)\approx 1/r$, and taking into account
that the original cell contains  $\pi L^2/(4\pi r_0^2)=N^{1/2}/4$
particles, we obtain (after integration) for the expectation value
of the  interaction energy per particle \be \dsf{E^{(I)}}{N}=
\dsf{a N^{1/2}e^2 \alpha^{1/2}}{L} \, . \lab{ap7} \ee Rewriting
the result for $E^{(I)}$ in $\hbar^2/(ML^2)$ units, substituting
into the expression for the expectation value of the total energy,
and minimizing with respect to $\alpha$ as in Ref. \ci{aormn} yields
the expression for the ground-state energy, Eq. \re{gsetup4}, with
$\hbar \omega_0=\hbar^2/(ML^2)$ and \be {\cal M}=\dsf{a
N^{1/2}L}{a_B} \lab{ap8} \, . \ee Following Ref. \ci{aormn} we assume
${\cal N}=1+\nu(N-1)$. The constant $a$ is fixed by fitting our
ground-state energy to the exact numerical value of the 2D Wigner
crystal ground state energy and we find $a = -c_{WC}$.

\section{Ground-state energy of 2$D$ Coulomb anyon gas}
\label{sec5}

For sufficiently large $N$ and by taking into account that $L=
N^{1/4} r_0$, we may use ${\cal M}=-c_{WC} N^{3/4}
r_s/f^{1/2}(\nu,r_s)$ and ${\cal N}=\nu N$ (in this section the
boson limit, $\nu = 0$, is understood as the limit $\nu
\rightarrow 0$ under the condition $\nu \gg 1/N$ ) to write,
starting from Eq. \re{gsetup4}, the expression for the ground-state 
energy per particle (in Ry units) in the form \be {\cal
E}_0(\nu, r_s)\approx
\dsf{2f(\nu,r_s)}{r_s^2}\left[\dsf{\nu}{2K_X^2}+\dsf{K_X^2}{2}-
\dsf{K}{K_X} \right] \, . \lab{gsqa11} \ee Here \be
K_X=(K_A+K_B)^{1/2}+[-(K_A+K_B)+2(K_A^2-K_A
K_B+K_B^2)^{1/2}]^{1/2} \, , \lab{gsqa6a} \ee and \bea \ba{l}
K_A=\left[K^2/128+\left((\nu/12)^3+(K^2/128)^2
\right)^{1/2} \right]^{1/3},\\
K_B=\left[K^2/128-\left((\nu/12)^3+(K^2/128)^2 \right)^{1/2}
\right]^{1/3}, \ \lab{gsqa9} \ea \eea with $K=c_{WC}
r_s/f^{1/2}(\nu,r_s)$, are obtained from Eqs. ~\re{gsetup5} and
\re{gsetup6} for $X_0$ and $A$ and $B$, respectively, after
substitution of  ${\cal M}$ and ${\cal N}$ by the expressions
given above. Note that we have scaled the lengths by
$f^{1/2}(\nu,r_s)$\,(see Sec. \ref{sec2}) and replaced $
\omega_0$ according to Eq. \re{gsetup12}. Finally, we determine the
function $f(\nu,r_s)$  by fitting  Eq. ~\re{gsqa11} to the known
analytic expressions for the ground state energy of spin-polarized
electrons and of the 2$D$ Coulomb Bose gas, as derived in Sec.
\ref{sec3}.

For the Bose gas ($\nu = 0$) we obtain from Eq. ~\re{gsqa11}  \be
{\cal E}_0(0, r_s)=-\dsf{c_{WC}^{2/3} f^{2/3}(0,r_s)}{r_s^{4/3}}
\lab{gsqa12} \ee and find with Eq. ~\re{2dbg8} for small $r_s$
$f(0,r_s)\approx c_{BG}^{3/2}r_s/c_{WC}$. For large $r_s$, the
ground-state energy does not depend on  statistics and equals the
energy of the classical 2$D$  Wigner  crystal \ci{bm},
$E_{WC}=-2.2122/r_s$. This matches with Eq.~\re{gsqa12} if at low
densities $f(0,r_s)\approx r_s^{1/2}$  with $c_{WC}^{2/3}=2.2122$. For
arbitrary $r_s$, we connect these asymptotics by \be f(0,r_s)\approx
\dsf{c_{BG}^{3/2}r_s/c_{WC}}{1+c_{BG}^{3/2}r_s^{1/2}/c_{WC}}+
\dsf{0.2 r_s^2 Ln(r_s)}{1+ r_s^2}\, . \lab{gsqa13}\ee

For $\nu\not=0$ and small $r_s$ the asymptotics of the ground-state 
energy, Eq.~\re{gsqa11}, has the form \be {\cal E}_0(\nu,
r_s \rightarrow 0)=\dsf{2 f(\nu,r_s)}{r_s^2}\left(\nu^{1/2}-
\dsf{c_{WC}r_s}{\nu^{1/4}f^{1/2}(\nu,r_s)}\right) \, ,
\lab{gsqa14} \ee the first term of which (for
$f(\nu,r_s)=1$) corresponds to the thermodynamic limit of the
ground-state energy of anyons in 2$D$ harmonic potential without
Coulomb interaction \ci{aormn,chitra}. For large $r_s$ we obtain \be
{\cal E}_0(\nu, r_s \rightarrow \infty ) =\dsf{c_{WC}^{2/3}
f^{2/3}(\nu,r_s)} {r_s^{4/3}}\left(-1+ \dsf{7\nu
f^{2/3}(\nu,r_s)}{3c_{WC}^{4/3}r_s^{4/3}}\right) \, . \lab{gsqa15}
\ee The function $f(\nu,r_s)$  has to fulfill the following
constraints: $f(\nu=1,r_s=0)=1$ for the dense ideal Fermi gas,
$f(\nu,r_s=0)=\nu ^{1/2}$ for the ideal anyon gas close to the
bosonic limit, and $f(0,r_s)$ given by  Eq. \re{gsqa13} for the
2$D$ Coulomb Bose gas. The interpolating functional form \be
f(\nu,r_s)\approx \nu^{1/2}c_0(r_s)e^{-5r_s}+
\dsf{c_{BG}^{3/2}r_s/c_{WC}}
{1+c_1(r_s)c_{BG}^{3/2}r_s^{1/2}/c_{WC}}+ \dsf{0.2 c_1(r_s)r_s^2
Ln(r_s)}{1+ r_s^2} \,  \lab{gsqa17} \ee with
$c_0(r_s)=1+6.9943r_s+22.4717r_s^2$ and $c_1(r_s)=1-e^{-r_s}$
satisfies these constraints and, in addition, yields in the
fermion case ($\nu=1$) for the ground-state energy per particle
the HF result \ci{rajagopal}. Moreover, for intermediate $r_s$ the
logarithmic term in Eq.~\re{gsqa17} gives ground-state energies of
the Bose gas lower than those of the spin-polarized electron
system, as indicated in \ci{DePalo}.

In Fig. 1 we show results for the ground-state energy per particle
on the large scale $1.0\leq r_s \leq15.0$. The upper four curves
refer in descending order to the fermion case ($\nu =1$): HF
energies for spin-polarized electrons from Ref. \ci{rajagopal} (open
triangles), interpolated by Pad$\grave e$ approximant MC data from Ref.
\ci{tanatar} (on the given scale identical with those of Ref. 
\ci{seidl}) for spin-polarized electrons (crosses), and our results
from  Eq. ~\re{gsqa11} (on the given scale identical with those of
Eq. ~\re{gsqa15}) for spin-polarized electrons (closed circles).
The lower two curves are for charged bosons ($\nu =0$) and result
from MC calculations of Ref. \ci{DePalo} (closed triangles) and from
our Eq. \re{gsqa12} (open squares). By star symbols we indicated
the MC data \ci{attac} (without interpolation) obtained for some
particular $r_s$ values. In Fig. 2 the corresponding results are
depicted for $0.2\leq r_s \leq 1.5$ and a larger energy scale.
Here the difference in the data from Eqs. ~\re{gsqa11} and
\re{gsqa15} (plus signs and closed circles, respectively) is
clearly resolved for the smallest $r_s$ values.

Let us first discuss the curves for the interacting Bose gas. As
can be seen from Figs. 1 and 2, our analytic formula \re{gsqa12}
for the ground-state energy yields results in close agreement with
the MC data of Ref. \ci{DePalo}, which are obtained numerically with
much effort and represent upper bounds to the ground state. In
this respect it is noteworthy that for $r_s>2.0$ the energies of
Eq. \re{gsqa12} are lower than those of Ref. \ci{DePalo}.

For the spin-polarized electron system we recognize, that the
results of our analytic formula (Eq. ~\re{gsqa11}) are lower than
the HF data for all $r_s$ with larger deviations for small $r_s$,
while the MC data (which are close to the HF results) correctly
approach the Wigner crystal limit (as our results). The curves
obtained from our low-density expansion, Eq. ~\re{gsqa15}, are
very close to those of Eq. ~\re{gsqa11} except for the lowest
$r_s$ values, thus indicating to breakdown of this approximation
(Fig. 2). The deviation from the HF and MC data for smaller $r_s$
can be made explicit by looking at the minima of these curves. By
minimizing Eq. ~\re{gsqa15} with respect to $f^{2/3}(\nu,r_s)/
r_s^{4/3}$ we obtain ${\cal E}_{0 \, , min}=-3c_{WC}^2/(28 \nu)$
for $\nu\not=0$. For the electron system ($\nu=1$) this gives with
$c_{WC}=3.2903$ the energy minimum ${\cal E}_{0 \, , min}=-1.159$.
With $f(\nu=1,r_s)$ from Eq.~\re{gsqa17} its position at
$r_s\approx 0.7$ is in close agreement with the result obtained
from Eq.~\re{gsqa11} (as shown in Fig.~2). In contrast, the HF
energy \ci{rajagopal} for the spin-polarized electron gas  \be
E_{HF}=\dsf{2}{r_s^2}- \dsf{16}{3\pi r_s} \, \lab{gsqa18} \ee
takes its minimum value $E_{HF\, , min}=-0.360$ for $r_s\approx
2.36$ compared with MC data of ${\cal E}_{MC \, , min}=-0.393$ at
$r_s\approx 2.3$.

Here we have to consider that the validity of the HF approximation
is limited by the demand that the leading (kinetic energy) is
larger than the second term (exchange contribution). For the
spin-polarized electron system this is the case for $r_s<1.2$,
i.e., the HF minimum energy per particle is achieved in a density
range, for which the HF approximation does not apply. Likewise, we
can discard the MC results for this intermediate density range
because they are obtained with a method, which conceptually is
close to the HF approximation.

Results for the $\nu$ dependence of the ground-state energy per particle,
calculated for various fixed values of $r_s$ using Eq.~\re{gsqa11}, are
presented in Fig. 3. It shows for $r_s > 1$ a monotonous increase from the boson
($\nu = 0$) to the fermion ($\nu = 1$) end. This follows analytically also by
taking the derivative of Eq.~\re{gsqa15}, valid for this range of $r_s$. In
contrast, for $r_s \leq 1$, a minimum appears for an intermediate value $\nu_0$,
thus the boson energies are not the lowest ones. This minimum follows (for $\nu
< r_s$) with $f(\nu,r_s\rightarrow 0)\approx\nu^{1/2}b_1$ with
$b_1=1+2.441472r_s$ from Eq.~\re{gsqa17} using the approximate Eq.~\re{gsqa15}.
It is given by ${\cal E}_{0 \, ,min}=-(4/5)b_2^{1/4}c_{WC}b_1^{1/2}/r_s$ with
$b_2=3/35$, and occurs at $\nu_0=b_2^{3/4}c_{WC}r_s/b_1^{1/2}$. For $r_s=0.2$
this gives ${\cal E}_{0 \,,min}\approx -8.689$ and $\nu_0\approx 0.08$, while
from Eq.~\re{gsqa11} we have ${\cal E}_{0 \, ,min}=-8.847$ and
$\nu_0=0.08$. From this analysis we see that for $r_s\rightarrow
0$ the minimal energy decreases according to ${\cal E}_{0 \, ,min}\sim -1/r_s$
much faster than the boson ground-state energy
with the asymptotics $\sim-1/r_s^{2/3}$. Therefore close to $r_s= 0$
the continuous transformation of bosons into fermions by varying
of parameter $\nu$ breaks down. This result is not surprising
because for $r_s\rightarrow 0$ we have no ideal boson gas.

Altogether the curves in Figs. 1 -- 3 demonstrate  the
capability of our analytical formula to describe the ground-state
energy in a wide range of the density parameter $r_s$ and for the
whole range of the anyon parameter $\nu$. The ground-state
energies for Coulomb anyon systems would continuously change
between the curves from Eq. \re{gsqa12} and Eqs. \re{gsqa11} and
\re{gsqa15} if $\nu$ would change from 0 to 1 (except the close region to $r_s=0$).

\begin{figure}
\begin{center}
\includegraphics[angle=270,width=14.5cm,scale=2.0]{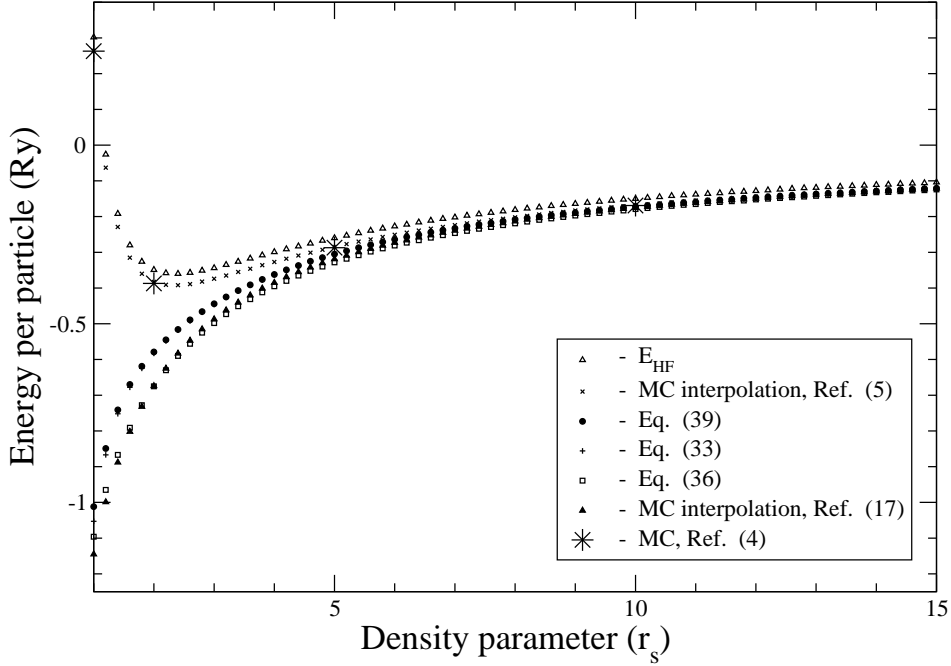}
\end{center}
\caption{Ground-state energies per particle vs density parameter
$r_s$ for range $1.0\leq r_s \leq 15.0$ from top to bottom: for
fermions ($\nu = 1$) HF approximation (Eq. ~\protect\re{gsqa18}
and Ref. ~\protect\ci{rajagopal}, open triangles), MC interpolation
data (from Ref. ~\protect\ci{tanatar}, crosses), and present results from
Eq. ~\protect\re{gsqa15} (closed circles) and Eq.
~\protect\re{gsqa11} (plus signs), and for bosons ($\nu = 0$) present
results from Eq. ~\protect\re{gsqa12} (open squares) and MC data
from Ref. ~\protect\ci{DePalo} (closed triangles). MC data of Ref. 
~\protect\ci{attac} for some particular values of $r_s$ are
indicated by star symbols.} \lab{fig1}
\end{figure}
\begin{figure}
\begin{center}
\includegraphics[angle=270,width=14.5cm,scale=2.0]{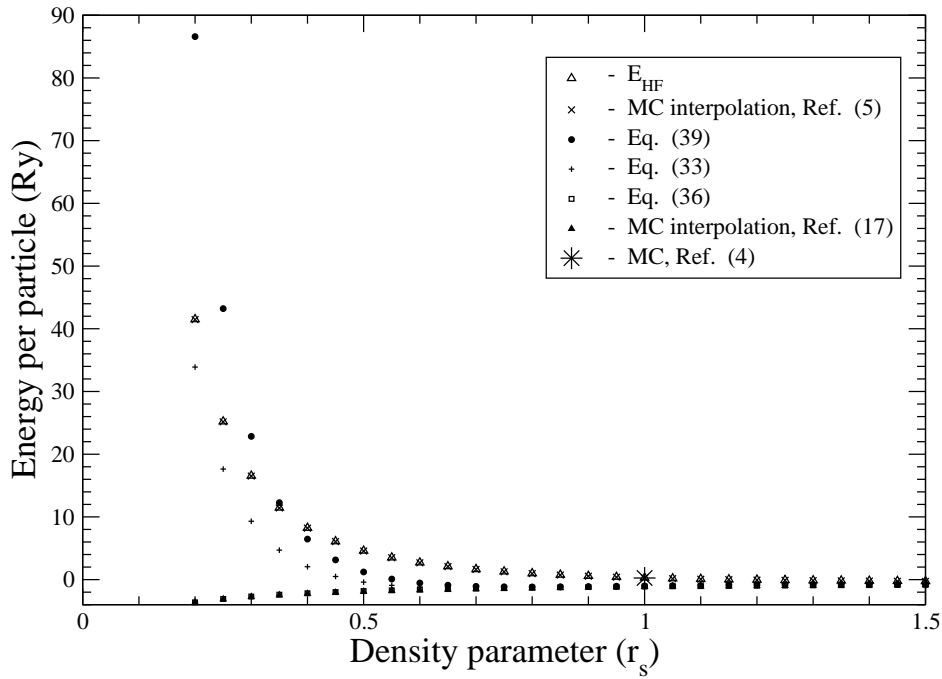}
\end{center}
\caption{ The same as in Fig. 1, however, for range $0.2\leq r_s
\leq 1.5$. } \lab{fig2}
\end{figure}
\begin{figure}
\begin{center}
\includegraphics[angle=0,width=14.5cm,scale=2.0]{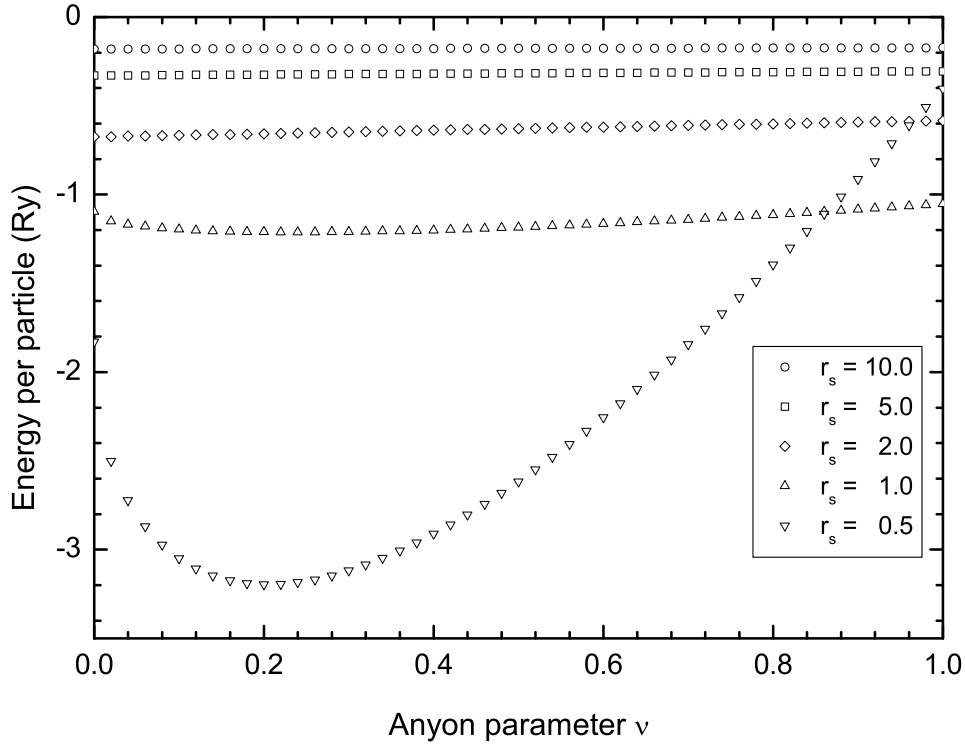}
\end{center}
\caption{ Ground-state energies per particle vs anyon parameter
$\nu$ for various fixed values of $r_s$.} \lab{fig3}
\end{figure}

\section{Conclusions}
\label{sec6}

We have derived an approximate analytic formula for the ground-state 
energy of the Coulomb anyon gas. Starting from our previous
results for the noninteracting 2$D$ anyon gas confined in a
harmonic potential, we have flattened out the confinement with
simultaneous increasing of the particle number to obtain the
thermodynamic limit at fixed density. We have generalized this
result to describe the interacting 2$D$ Coulomb anyon gas by
introducing a function of the anyon parameters $\nu$ and $r_s$,
that takes into account the Coulomb interaction and fractional
statistics. We have determined this function by fitting to the
analytic expressions for the ground-state energy of the classical
electron crystal at very large $r_s$, to that of the 2$D$ Coulomb
Bose gas at very small $r_s$ and to the HF energy (at high
density) for spin-polarized 2$D$ electrons. Our analytic formula
applies to the full range of parameters $\nu$ and $r_s$ and
provides a convenient description of the ground-state of 2$D$
Coulomb anyon systems. By comparison with HF and MC results from
the literature we find significant deviations for intermediate
densities, which indicate shortcomings of these approaches.

\section{Acknowledgments}

B. A. and M. M. acknowledge support received from the Volkswagen
Foundation and the hospitality at the University of Regensburg. We
thank G. Ortiz for the idea of harmonic potential regularization
to obtain the thermodynamic limit and M. Seidl for stimulating
discussions.

\end{document}